\documentclass[pra,twocolumn,letterpaper]{revtex4}

\usepackage{graphicx}
\usepackage{hyperref}
\usepackage{amsthm}
\usepackage{amsmath}
\usepackage{amsfonts}
\usepackage{verbatim}

\begin{document}

\preprint{PRE/003}

\title{Efficient Heralding of Photonic Qubits  with Applications to Device Independent Quantum Key Distribution}

\author{David Pitkanen$^{1}$, Xiongfeng Ma$^{1}$, Ricardo Wickert$^{2}$, Peter van Loock$^{2}$, Norbert L{\"u}tkenhaus$^{1}$}
\affiliation{$^1$Institute for Quantum Computing \& Department of Physics
  and Astronomy, University of Waterloo, 200 University 
Avenue West, N2L 3G1 Waterloo, Ontario, Canada}
\affiliation{$^2$Max Planck Institute for the Science of Light and Universit\"at Erlangen-N\"urnberg, 91058 Erlangen, Germany}

\date{\today}


\begin{abstract}
We present an efficient way of heralding  photonic qubit signals using linear optics devices. First we show that one can obtain asymptotically  perfect heralding and unit success probability with growing resources. Second, we show that even using finite resources, we can improve qualitatively and quantitatively over earlier heralding results. In the latte r scenario, we can obtain perfect heralded photonic qubits while maintaining a finite success probability. We demonstrate the advantage of our heralding scheme by predicting key rates for device independent quantum key distribution, taking imperfections of sources and detectors into account. 
\end{abstract}

\maketitle

\section{Introduction}
Applications such as the verification of entanglement for quantum communication, and the establishment of security proofs in Device Independent Quantum Key Distribution (DIQKD) \cite{Pironio2009a} have generated increasing interest in violations of Bell's inequalities over long distances. A violation of Bell's inequality is supposed to be a test for a non-local correlation between the outcomes of pairs of events; however, it is difficult to design an experiment that rigorously shows this non-locality, as imperfections in the experimental equipment can open loopholes which allow for a local Hidden Variable (HV) explanation of the measured data. For example, signal loss in optical implementations generates the so-called {\em detection efficiency loophole} \cite{Pearle1970}.

In experiments, with growing distance the transmission loss increases. Consequently, the resulting low total detection probabilities make violations of Bell's inequality virtually impossible. To overcome this issue, the use of heralding devices \cite{RalphLund08,gisin10a} has been suggested. Such an apparatus performs a measurement that ressembles a quantum non-demolition (QND) measurement, raising a flag to indicate whenever the desired signal successfully traversed the channel. The state generated by conditioning on this flag can then be employed in a Bell test. This procedure does not lead to a detection loophole as long as the flagging is independent of the measurement choice.

Gisin et al.~\cite{gisin10a} have considered an implementation of a heralding device in a DIQKD scheme employing realistic sources, linear optical components and photon-number resolving detectors. The proposed scheme uses the transmissivity of one of its beam splitters as an adjustable parameter to regulate the ratio between vacuum and single photons in the conditional output state at the expense of the probability of successful heralding. With input signals consisting only of vacuum and single-photon states, increasing the single-photon component to unity in the conditional state can only be achieved in the limit of vanishing success probability. If the input also contains multiphoton signals, then the fraction of single photon signals in the conditional states cannot reach unity. 

In the present work we investigate improvements on this scheme to overcome its limitations. We begin by reviewing the known results on heralding in Section \ref{sec:photonic}. Then in Section \ref{sec:klm}, we investigate  fundamental limits with unlimited resources, staying within the linear optics toolbox. In this regime, we can employ large ancilla states to realize a KLM-like teleportation procedure \cite{knill01a} and demonstrate that we can perform the desired QND measurement perfectly, although non-deteministically. Shifting our focus to practical schemes that allow only limited resources, we then revisit in Section \ref{sec:modified} the original proposal. Motivated by our general scheme and the Hong-Ou-Mandel effect \cite{Hong87}, we suggest a modification that significantly improves the heralding device. The modification is done by adding two beamsplitters to the initial linear optics circuit while maintaining the original simple ancillae states.

Since the original amplifier was proposed for use in DIQKD, we analyze the performance of our amplifier in Section \ref{sec:method3} and  demonstrate that the improvement of the heralding device translates into an enhanced key rate.

\section{Photonic and Qubit Amplifier}
\label{sec:photonic}
\begin{figure}
\includegraphics[trim = 1cm 18cm 1cm 0cm, clip, scale=0.4]{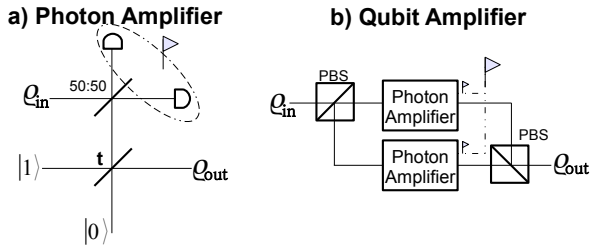}
\caption{\label{amplifier}
(a) The Ralph-Lund noiseless linear amplifier scheme: an input mode $\rho_{in}$ interacts with an ancilla state through a $50:50$ beam splitter. Conditioned on a successful detection pattern, which raises the heralding flag, the output $\rho_{out}$ is shifted towards the single-photon state. The parameter $t$ is the transmissivity of the beamsplitter (b) The Gisin-Pironio-Sangouard qubit heralding device: two amplifiers are combined to amplify states in the horizontal/vertical ($h/v$) basis; the flag is only raised if both of the amplifiers are successful. The input state $\rho_{in}$, encoded in the polarisation basis $h/v$, is sent through a polarising beamsplitter (PBS) to spatially separate its modes so that the different amplifiers may be applied.  A second PBS is used to combine the different spatial modes of the output into the $h/v$ basis.  In both schemes the feed-forward mechanism has been omitted.}
\end{figure}

We begin by revisiting the noiseless linear photonic amplifier proposal of Ralph and Lund \cite{RalphLund08}, see Fig.~\ref{amplifier}. A single-photon ancilla state passes through a beam splitter of transmissivity $t$ to create an entangled state of two modes involving vacuum and single photons. One of the modes will be the output of the device, while the other is mixed with the input mode, $\rho_{in}$, on a 50:50 beam splitter. Both output modes of the $50:50$ beam splitter are measured on photon detectors and the observation of exactly one photon in the measured modes is taken as the successful heralding flag. 

Depending which of the two detectors is triggering the flag, an optical phase correction has to be applied at the output of the device. This feed-forward mechanism is not essential to our discussions, and we incorporate it directly into the description of the device. In practise one will have to do this feed-forward, unless the action of the phase correction can be combined with a subsequent measurement in such a way that, instead of active feed-forward, just a re-interpretation of the measurement results takes place. 

For $t=\frac{1}{2}$, this scheme amounts to standard probabilistic quantum teleportation; however, for $t>\frac{1}{2}$, the vacuum component $|0\rangle$ of the outgoing mode $\rho_{out}$ is reduced and the single-photon term $|1\rangle$ emerges enhanced relative to the vacuum component. This corresponds to a mapping induced by a Kraus operator $A$,
\begin{align}
A \left(c_{0}|0\rangle+c_{1} |1\rangle \right) =\sqrt{1-t}c_{0}|0\rangle+\sqrt{t} c_{1} |1\rangle \; ,
\end{align}
which has the important property that, as $t \rightarrow 1$, this circuit approaches a projection onto the single photon state, $|1\rangle$.  
We use non-normalized states in our description, so that the success probability of the heralding device is given by the norm of the conditional output state and it is naturally dependent of the input state.

The two-dimensional Hilbert space of exactly one excitation in two optical modes is known as a photonic or dual-rail qubit \cite{kok07}. It is the central idea of the work by Gisin et al.~\cite{gisin10a} to use two Ralph-Lund amplifiers \cite{RalphLund08} in parallel on the two modes to herald  such a photonic  qubit. A successfully heralded event  is defined as the joint success of both amplifiers. The action is then represented as 
\begin{multline}
A \otimes A \left( c_{00} |00\rangle + c_{10} |10\rangle + c_{01} |01\rangle \right) = \\ 
(1-t) c_{00}|00\rangle
+\sqrt{t(1-t)} \left(  c_{10} |10\rangle + c_{01} |01\rangle \right)\; .
\label{inState}
\end{multline}
The circuit for this qubit amplifier is shown in Fig. \ref{amplifier}, where the dual-rail qubit is encoded in the polarization. The weight of the dual-rail qubit component in this conditional state can reach unity in the  limit $ t \rightarrow 1 $. In this limit  the probability of successful heralding vanishes, indepedendently of the input state. If the input state also contains a  multiphoton component, $ c_{11}|11\rangle$, then the weight of the dual-rail qubit in the  output state can no longer reach unity, as the output will also contain the component $t c_{11}|11\rangle$. In that case, the choice $\frac{t}{1-t} = \frac{|c_{00}|}{|c_{11}|} $ optimizes the qubit fraction of the output.  This optimal qubit fraction in the heralded signals is then given by
\begin{equation}  \frac{|c_{10}|^2 + |c_{01}|^2}{2 |c_{11}| |c_{00}|+|c_{10}|^2 + |c_{01}|^2} \; .\label{sphotonbound}
\end{equation}
Although we have only considered the case of a pure state input, this bound also applies to mixed states if $\sqrt{\langle ij | \rho_{in}|ij \rangle}$ is used instead of $|c_{ij}|$.

\section{KLM Procedure}
\label{sec:klm}
Before proposing a scheme to overcome the limitations of the heralding setup by Gisin et al., let us take a more fundamental point of view. We remain restricted to the linear optics toolbox, but allow for the use of more complicated sources for the ancilla states, and show that, in this context, the heralding measurement for dual-rail qubits can be performed asymptotically perfectly. Our approach is based on the KLM framework \cite{knill01a} and the procedure introduced therein, which successfully accomplishes the teleportation of an arbitrary state of the form $c_0 |0\rangle + c_1 |1\rangle$ with a probability that can be brought asymptotically close to $1$.

The implementation of such procedure relies on the use of an ancilla state
\begin{eqnarray}
\label{tnorig}
|t_n \rangle &=& \frac{1}{\sqrt{n+1}}\sum_{i=0}^n  |s_{n,i}\rangle \; ,
\end{eqnarray}
where $|s_{n,i}\rangle = |1\rangle^i|0\rangle^{n-i} |0\rangle^{i} |1\rangle^{n-i}$. Here the notation $|1\rangle^i$ refers to the $i$-fold tensor product $|1\rangle^{\otimes i}$. We refer to the first $n$ modes of this state as the teleporting modes, while the second $n$ modes are referred to as output modes. In order to perform the teleportation, a $(n+1)$-mode Fourier transform is performed on the teleporting modes of the ancilla $|t_n \rangle$ and the input, followed by a photon-counting measurement on these modes. The Fourier transform is responsible for making the result of the photon measurements indistinguishable independently of which modes the photons have originated from. As a result, if $k$ photons are detected, the remaining unmeasured output modes are left in the state
\begin{align}
c_0|0\rangle^{k} |1\rangle^{n-k}+e^{-i\phi_k} c_1|0\rangle^{k-1} |1\rangle^{n-k+1} \; .
\end{align}
The phase $\phi_k$ depends on the number of counted photons, $k$, and the observed  detection pattern, and can be corrected with an appropriately adjusted phase shifter.  As in the case of the linear photonic amplifier, we will incorporate this correction automatically into our description. Overall, the input state will then be found in the $k$-th mode of the above state.

This approach can also be adapted to perform a non-demolition measurement onto the single-photon input space of two modes. To do this, we will consider the auxiliary state
\begin{equation}
\label{tntilde}
|\tilde{t}_n \rangle = \frac{1}{\sqrt{n+1}} \sum_{i=0}^n|s_{n,i} \rangle|s_{n,n-i}\rangle \; ,
\end{equation}
which corresponds to those terms in $|t_n\rangle \otimes |t_n \rangle$ containing exactly $n$ photons in the two sets of teleporting  modes in the $|s_{n,i}\rangle$ states. We now show that the application of  the KLM-type procedure, employing the alternative state above, realizes a QND measurement  onto the total photon-number space of the input modes. To verify this, we first note that this procedure effectively measures the photon number in the input: as the total number of photons in the two pairs of teleporting auxilliary modes is known to be $n$, the observed photon number on the input and these $2 n$ modes tells us how many photons have entered the heralding device. In a second step, we need to verify that the output state corresponds to that of a QND measurement: if the two Fourier measurements acting each on one input mode and one set of teleporting modes yield the observation of $i$ and $n-i+1$ photons  respectively, giving exactly $n+1$ photons in total,  and these individual photon numbers are neither $0$ or $n+1$, then the corresponding conditional state of the remaining  $2 n$ output modes is
\begin{align}
c_{01} |0\rangle^i |1\rangle^{n-i} |0\rangle^{n-i} |1\rangle^{i}+c_{10}|0\rangle^{i-1} |1\rangle^{n-i+1} |0\rangle^{n-i+1} |1\rangle^{i-1} \; .
\end{align}
This means that the input state has been teleported into the mode pair with indices $(i,2n-i+1)$ of the above state. The probability of failure of this scheme, just as in the original KLM proposal, is connected to occurence of $0$ or $n+1$ photons in the individual Fourier measurements, and is given by $\frac{1}{n+1}$. Thus, one can (asymptotically) not only perform a perfect heralding measurement, \textit{i.e.}, approaching the idealized QND measurement in the desired single-excitation subspace, but also have the probability of success be made arbitrarily close to unity.

\section{Modified Amplifier Circuit}
\label{sec:modified}
An obvious strength of the scheme proposed in \cite{gisin10a} lies in the relative simplicity of its ancilla states, which can be generated with a single-photon source and vacuum states. Here we take a practical approach, keeping the same ancilla states, and look for simple modifications  we can make to the amplifier to improve its performance. We focus on modifying the amplifier so that a vacuum input will no longer trigger the heralding flag at all.
We begin by examining the original scheme, which consists of two separate Ralph-Lund amplifiers, each with their own auxilliary single photon states.

In order for the signal to be heralded by the qubit amplifier, both of the flags on the separate Ralph-Lund amplifiers need to be raised by detecting  exactly one photon respectively, after each of the 50:50 beam splitters. This set-up can lead to false flagging for a vacuum input. These false heralding flags occur if  both of the auxiliary photons from the separate  amplifiers  travel 'upwards' in our diagram towards the heralding detectors behind the 50:50 beamsplitters.

\begin{figure}[h!]
\includegraphics[scale=1]{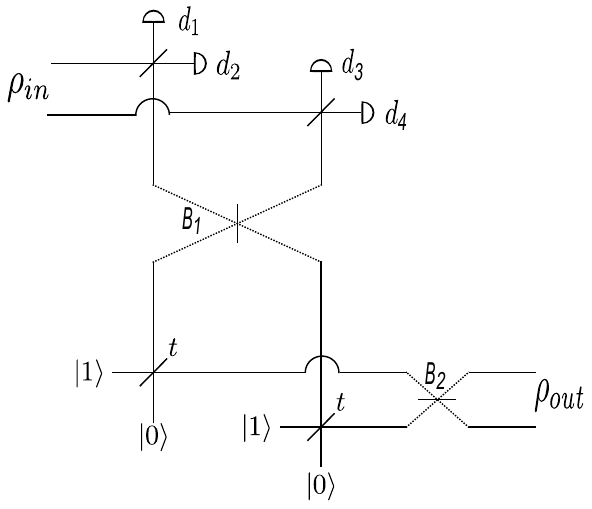}
\caption{\label{circuit} The proposed  circuit for the improved qubit amplifier.  Without the two $50:50$ beamsplitters, marked $B_1$ and $B_2$, it corresponds exactly to  exactly the amplifier suggested by Gisin et al.  The circuit's inputs are the two modes of $\rho_{in}$.  The amplifier is only meant to output a signal, $\rho_{out}$, when a single photon is measured in each of the detector sets  $(d_1,d_2)$ and $(d_3,d_4)$. The required feed-forward mechanism to correct optical phases is omitted. 
}
\end{figure}

We suppress this component by adding another 50:50 beamsplitter between these upward directed modes; the Hong-Ou-Mandel effect \cite{Hong87} ensures that the component with two photons in each mode will now bunch to either of the outgoing modes of the beamsplitter. This means that the heralding detectors of one of the Ralph-Lund amplifiers will see zero photons, while the other will see two, and therefore, the heralding condition of the qubit amplifier will no longer be met.
A second 50:50 beamsplitter is added to the output modes of the qubit amplifier, so that the transformation effected by the amplifier to the single-photon input does not change. On the other hand, the action of the second beamsplitter corresponds to a change of polarization basis on the single-photon subspace, and can be absorbed into the action of any device that is acting on the output of the heralding device. With these two additional beamsplitters  (see Fig. \ref{circuit}), we find that the successful heralding is connected to the Kraus operator $A_{mod}$ given by

\begin{eqnarray}
\lefteqn{A_{mod}\left( c_{00} |00\rangle + c_{10}|10\rangle + c_{01} |01 \rangle + c_{11} |11\rangle \right)}& &  \\ 
&=& \sqrt{t(1-t)}  \left( c_{01}| 01 \rangle
+ c_{10}| 10 \rangle \right) +
t\frac{1}{\sqrt{2}} c_{11}\left( | 20 \rangle + | 02 \rangle \right)\; . \nonumber
\end{eqnarray}
This Kraus operator already accounts for the required phase correction that depends on the exact pattern of single-photon detection after each of the two 50:50 beramsplitters.

The transformation overcomes both  problems that were discussed in the previous  section: first, assuming only vacuum and single photon input signals ($c_{11}=0$), this amplifier can perform a perfect, heralded projection onto the dual-rail component.  Second, even if multiple photons are present  in the input ($c_{11} \neq 0$), the single-photon fraction in the output can still be made arbitrarily close to unity in the limit of vanishing success probability ($t \longrightarrow 0$) .

\section{Application to Device Independent QKD}
\label{sec:method3}
The security of Device Independent QKD (DIQKD) is based on a set of assumptions that is reduced compared to that of traditional QKD; the most remarkable difference being that for DIQKD the communicating users of an entanglement-based scheme  can be ignorant of the precise characterization  of their measuring devices without compromising the security of the protocol.  The privacy of the resulting key depends on the users being able to perform measurements that violate Bell's inequality. This approach first appeared with EPR based QKD \cite{ekert91a}. Related work has been done under the headline of self-testing devices \cite{mayers98suba,Mayers04a,McKague10}. Here we refer to the formulation by Acin et  al. \cite{acin07a}.

To bound information of an eavesdropper in DIQKD, Alice and Bob both randomly and independently select two measurements which are designed to verify the violation of Bell's inequality by evaluating a Bell parameter $S$.  To generate the key, the receiver also makes use of a third measurement, $\sigma_z$, which is designed to obtain highly correlated data with the sender.  These data serve as the raw key from which the final key will be distilled. On these data, we expect to find binary error rates $Q$.

In the proposal for the original qubit heralding device, a DIQKD simulation was performed to demonstrate how heralding overcomes transmission losses. The simulation included imperfect sources and detectors. We perform analogous simulations to demonstrate the improvement that our heralding device offers. For this comparison, we consider three main scenarios. The first one is motivated by the simulation reported in  Gisin et al. \cite{gisin10a}, where the authors introduced a theoretical framework to deal with inconclusive outputs due to imperfect devices. In this framework restrictions on the eavesdropping strategies are assumed; we therefore refer to this framework in our simulations as \textit{Restricted Device Independent Theory}, in subsection \ref{ResDeInThe}. In addition we also run simulations which deal with the inconclusive results by randomly assigning them a conclusive binary value, thereby allowing us to apply the usual  \textit{Unrestricted Device Independent Theory}, in subsection \ref{UnrDevIndThe}. As a third framework, we explore the so-called \textit{Detection Device Independent Theory}, in subsection \ref{DetDevIndThe}, where knowledge of the source is assumed, but one can remain ignorant about the measurement device of one of the parties. The assumed knowledge of the source, in this framework, makes it useful not only for entanglement-based setups, but also for prepare-and-measure schemes.

\subsection{Experimental Setup}
\begin{figure*}
\includegraphics[trim = 0cm 21cm 0cm 0cm, clip, scale=0.75]{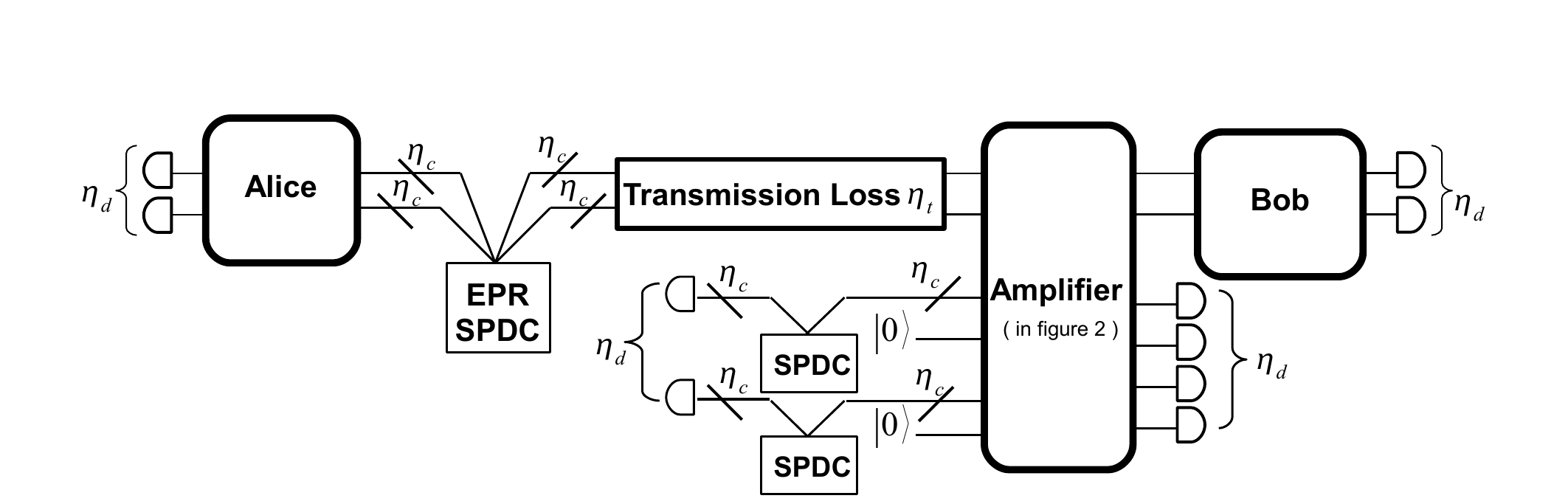}
\caption{\label{setup}
Experimental setup for the amplifier in the DIQKD simulations we consider.  An EPR SPDC source is used to generate photons that are sent to the two distant parties. Two additional SPDC sources are used as heralded single photon sources for the amplifier.  Each wire in this diagram represents a separate optical mode --- ie. we use two wires to represent a single spatial mode if we are using both of the polarisation, horizontal/vertical $(h/v)$, degrees of freedom.  The signal recieved by Bob is only processed if the right detection pattern appears on the amplifier.
}
\end{figure*}

We now describe the proposed experimental setup for DIQKD.  To mediate the communication, a spontaneous Parametric Down-conversion Source (EPR-SPDC) is used  which generates entangled photons. The (unnormalized) state obtained through this process is given by
\begin{align}
\rho_{source} = |0 \rangle\langle 0| + p | \phi^{+} \rangle\langle \phi^{+}| + p^{2} |\phi^{+2}\rangle\langle\phi^{+2}| + O(p^3) \; ,
\end{align}
where $|\phi^{+}\rangle$ is an EPR pair, $|1010\rangle+|0101\rangle$; and $|\phi^{+2}\rangle = |2020\rangle + |1111\rangle + |0202\rangle$, up to a normalization. The parameter $p$ is related to the pumping power.
This EPR-SPDC source is located near one of the parties, Alice, and the two-mode signal that is received by the more distant party, Bob, is subject to transmission loss $\eta_t$. The loss which results from using imperfect detectors and coupling into fibers is taken into account with efficiency parameters $\eta_d$ and $\eta_c$ respectively. In our simulations we assume that all the detectors have the same efficiency so we model the detectors as perfect and include the detector loss in the coupling efficiency $ \eta_{cd} = \eta_c \eta_d $.  The photons employed in the auxilliary states of the amplifiers are generated from a heralded SPDC process that outputs the state \cite{Pittman05}
\begin{eqnarray}
 \rho_{aux} &= &p' \eta_{cd} |1\rangle \langle 1| + 2 (1-\eta_{cd}) \eta_{cd} p'^2 |2\rangle \langle 2|+ \nonumber\\
& & 3 (1-\eta_{cd})^2  \eta_{cd} p'^3 |3\rangle \langle 3| + O(p^4) \; ,
\end{eqnarray}
where $p'$ is again the pumping power. 
The amplifier therefore acts on the state
\begin{align}
\rho_{total}= \rho_{source} \otimes {\rho_{aux}^{\otimes 2}} \otimes |0 \rangle \langle 0 |^{\otimes 2}
\end{align}
which incluces both components of the source states: the one that remains at Alice's site, and that which enters the amplifier. The whole set-up is depicted in Fig.~\ref{setup} to indicate  where coupling and detection efficiencies are included. Note that we include coupling efficiencies for the heralding detectors .  Omitting these reproduces for the original heralding device exacltly the simulation shown in \cite{gisin10a}. The source is on Alice's side of the set-up. Therefore, transmission loss affects only the signal travelling from the source to the amplifier, which is located in Bob's site. 

For simulation purposes, we follow the choices made in \cite{gisin10a} and assume a repetition rate of $10$ GHz.  In order to maximize the key rate, an optimization is performed over the pump parameter $p$  and over the transmissivity $t$ of the beamsplitter used in the heralding device. The range of the parameters $p$ and $p'$ are restricted to $0 \leq p,p' \leq 10^{-2}$, as we use a perturbative approach in our simulations. This constraint, however, affects only the simulations of the Detection Device Independent Scenario.

The detectors in the heralding schemes are modelled as photon-number resolving detectors. Transition-edge superconducting detectors can be used for this purpose. Alternatively, a cascade of beamsplitters followed by threshold photon counters can approximate such photon-number resolving detectors. 

Our calculations do not include dark counts. For this to be a valid assumption, we will require that the total dark count rate is negligibe compared to the  total rate of heralded events. Superconducting nanowire single-photon detectors have been demonstrated to be able to work in this regime \cite{Lita08}.

Our simulations are done as perturbative {\em approximations} in the pump parameters $p$ and $p'$. To bound the error in this approximation, we also provide {\em lower bounds}  on the expected key rates by calculating the total weight of the neglected terms, and by using this weight in independent worst case values for the Bell parameters $S$ and quantum bit error rates $Q$.

\subsubsection{Restricted Device Independent Theory}
\label{ResDeInThe}
\begin{figure}[h!]
\includegraphics[trim = 2cm 7.6cm 2.5cm 8cm, clip, scale=0.5]{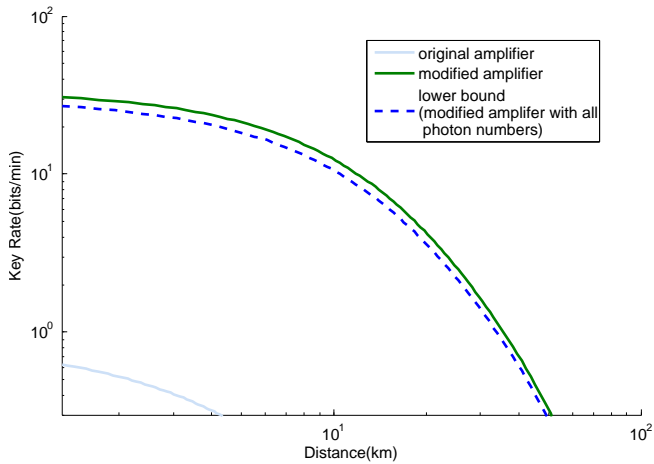}
\caption{\label{PartSec} Key Rate vs. Distance, plotted for both the original \cite{gisin10a} and the modified amplifier.  The simulations are done using the \textit{Restricted Device Independent Theory} framework. The key rate is calculated from Eqn.(\ref{keyRest}), multiplied by the repetition rate of the source and the probability that the amplifier successfully heralds the signal.  The efficiency parameters are chosen as  $\eta_d =0.95$ and $\eta_c=0.90$, resulting in an overall efficiency of $\eta_{cd} = 0.855$. }
\end{figure}

The framework proposed by Gisin \textit{et al.} follows the standard device independent protocol \cite{Acin06}, but augments it by an analysis which makes an additional assumption about the eavesdropping strategies. Thanks to this assumption,  all of the inconclusive results can be discarded during post-processing, though the rate of inconclusive results affects the resulting key rate. The key rate, which is given by

\begin{eqnarray}
K \ge  & & \mu_{cc}  \bigg\{  1- h[Q_{cc}]  \label{keyRest} \bigg. \\
&  & \left. - \left[ \left( 1-\frac{\mu_c}{\mu_{cc}} \right) \chi \left[\frac{\mu_{cc} S_{cc}-4\mu_c}{\mu_{cc}+\mu_c}\right] + \frac{\mu_c}{\mu_{cc}} \right] \right\} \; ,\nonumber
\end{eqnarray}
with
\begin{eqnarray}
h[x] &=& -x\log_2[x] - (1-x) \log_2{[1-x]}   \; \mbox{ and}\label{entropy}  \\
\chi[x] &=& h\left[ \frac{1+\sqrt{(x/2)^2 -1}}{2} \right]  \label{chi} \; .
\end{eqnarray}
Here, $\mu_{cc}$ is the probability that both parties obtain a conclusive result. Within this set of conclusive data,  $S_{cc}$ is the measured Bell parameter, and $Q_{cc}$ is the error rate. The probability that only one of the parties obtains a conclusive result is denoted by $\mu_c$.  The results are shown in Fig. \ref{PartSec}.  We find that our heralding device improves the distance and rate significantly.
   
\subsubsection{Unrestricted Device Independent Theory}
\label{UnrDevIndThe}
The scheme utilizing the unrestricted device independent theory differs solely from the setting in the previous section in its data post-processing stage. Any inconclusive measurement result on either side has binary outcomes assigned at random. Knowledge of the placements of such inconclusive results is later used in the error correction step \cite{Ma11}. For this reason, the key rate includes quantities that reflect the fraction of conclusive results. The resulting key rate is

\begin{equation} K = \mu_{-c}\left(1-h[Q_{-c}]\right) - \chi[S] \; .\label{XFkRate}
\end{equation}
Here, $\mu_{-c}$ is the probability of Bob obtaining a conclusive result and $Q_{-c}$ is the error rate within  Bob's conclusive measurement results. Finally, $h[x]$ and $\chi[x]$ are the same as in Eq.'s (\ref{entropy}) and (\ref{chi}).  The  Bell parameter $S$ is evaluated using the data from all of the measurement results, including the  the random assignments of inconclusive results.

From Eq.(\ref{XFkRate}), we can see that the parties need non-classical correlations with  $S>2$ to generate a positive key.  Therefore, the random assignment puts a constraint on the probability that a conclusive result is measured on both sides, $\mu_{-c} > \frac{1}{\sqrt{2}} \sim 0.707$.  However, if we use a SPDC source to generate the signals, we find from Eq. (\ref{sphotonbound}) that the qubit fraction after heralding is bounded by $\mu_{-c} < 3/5 $ when using the original heralding device, even when using ideal detectors and single photon sources. Therefore, such amplifier cannot be employed to generate positive key rates in this framework, unless a different source is used to generate the entangled photons. Note that other assignments of inconclusive results are possible \cite{Pironio2009a} which may allow to extract secret key with the original heralding device. The discussion requires more detailed analysis, as it depends on the exact configuration of the set-up. It is omitted here, as these studies go beyond the scope of the current research. However, some  discussion of this question can be found in the work by Moroder and Curty \cite{curty11a}. 

Due to the above, the simulations for the unrestricted device independent theory are performed only for the newly proposed amplifier (see Fig \ref{FullSec}).  This framework is the most demanding on coupling and detection efficiencies that need to be used in order to generate a positive key: in our simulations, we use a total loss term $\eta_{cd} = 0.93 $.
\begin{figure}[h!]
\includegraphics[trim = 2cm 8cm 3cm 8cm, clip, scale=0.5]{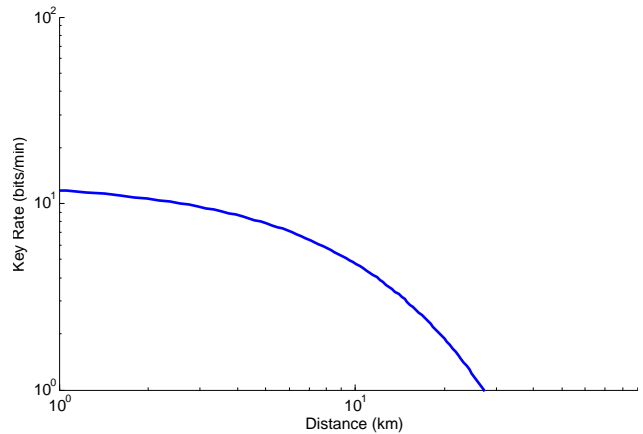}
\caption{\label{FullSec} Key Rate vs Distance, plotted on a logarithmic scale for our proposed amplifier using the \textit{Unrestricted Device Independent Theory} framework. We calculate the key rate from Eq. (\ref{XFkRate}), multiplied by the repetition rate of the source, $10$ GHz, and the probability that the amplifier successfully heralds the signal. The efficiency parameter is $\eta_{cd}=0.95$. 
}
\end{figure}

\subsubsection{Detector Device Independent Theory}
\label{DetDevIndThe}
The final framework we consider is not fully device independent and, as a result, the security does not require a loophole-free violation of Bell's inequality. Here, the standard BB84 protocol is used \cite{bennett84a} and we trust the source on Alice's side only. Bob's detectors remain uncharacterized.  This scenario has been considered by Mayers \cite{mayers01a}, and later also by Koashi.
The scenario makes random assignment of inconclusive results on Bob's side necessary and therefore places  a constraint on the detection probability that is required in order to generate a secure key \cite{Ma11}. Again we use the fact that the position of events with random assignment are known to Bob, who can utilize this knowledge in the later error correction step. The security proof \cite{Ma11} is therefore a variation of corresponding proofs of Koashi \cite{koashi2009}.
This scenario is more tolerant to transmission loss. For example, with perfect photon pair sources and detection devices, this scenario tolerates a total efficiency accounting for transmission, coupling and detection loss of $64.5\%$ without heralding \cite{Ma11}. The key rate this scheme is given by \cite{Ma11}
\begin{align}
K \ge \mu_{-c}\left(1-h[Q_{-c}]\right) - h[\delta_b] \; .\label{detIndK}
\end{align}
Here, $\delta_b = \mu_{-c} Q_{-c} + (1-\mu_{-c}) \frac{1}{2}$ is an effective phase error rate; as in Eq.(\ref{XFkRate}), $\mu_{-c}$ is the probability of Bob obtaining a conclusive result, and $Q_{-c}$ is the error rate when Bob's measurement is conclusive. This framework is the least demanding on the the coupling and detector efficiencies. 

The simulations' results shown in Fig.~\ref{FullSec}.  In this case the pump parameters are larger compared to the other two scenarios.  Therefore, the gap between the approximated rates and the lower bounds is more pronounced.
\begin{figure}[h!]
\includegraphics[trim = 1.4cm 8cm 3cm 8cm, clip, scale=0.5]{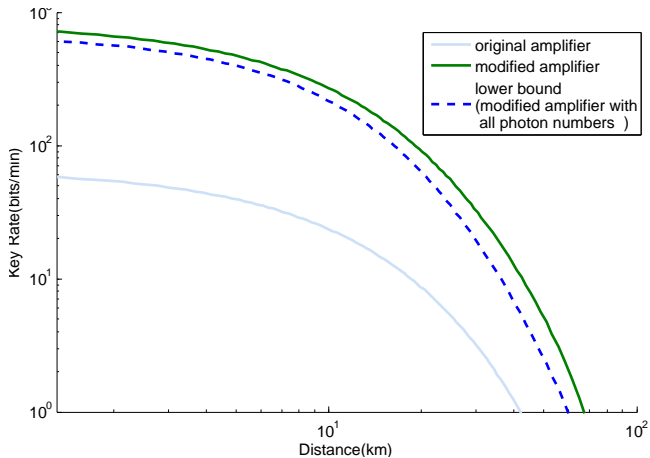}
\caption{\label{FullSec} 
Key Rate vs Distance, plotted for both the original and the modified amplifier using the \textit{Detection Device Independent Theory} framework. We calculate our key rate from Eq. (\ref{detIndK}), multiplied by the repetition rate of the source, $10$ GHz, and the probability that the amplifier successfully heralds the signal. The efficiency parameter is $\eta_{cd} =0.75$.}
\end{figure}

\section{Conclusions}
Heralding devices can play an important role in quantum key distribution. In principle, they allow to overcome the limitation posed by transmission losses to device independent quantum key distribution. In addition, other areas of quantum communication can benefit from such heralding devices. For example, some quantum memory approaches do not provide intrinsic heralding devices. Using external heralding, as proposed in this paper, will allow the use of such memories in  quantum repeater technologies \cite{sangouard11}. 

In this regard, we have explored the KLM framework and employed it in the implementation of a conceptually optimal heralding strategy which, in the limit of asymptotic resources, achieves a perfect QND-like measurement onto the desired signal subspace with success probability approaching unity.

Departing from the conceptual scenario, we discussed a simple and experimentally-viable improvement on the original work by Gisin \textit{et al.} \cite{gisin10a}, enhancing the performance of the heralding device. Specifically, we were able to overcome the undesired relation between how reliably the device works in heralding its input, and the success probability of the heralding process. Our device allows, in an idealistic implementation, perfect operation on the important input subspace containing at most one photon in each of the optical modes that define the dual-rail qubit. 

This improvement not only increases the achievable key rates, in the context of a restricted device independent theory, by roughly one order of magnitude in distance and rate as compared to the scheme in \cite{gisin10a}, but also allows us to enter the domain of fully unrestricted device independent theory. Our simulations show that, under the similar assumptions as those made in \cite{gisin10a}, we can obtain positive key rates in this desirable scenario. Note, however, that the requirements on detection and coupling efficiency are  more demanding. 

Finally, we showed that, for detector device independent theories with a well characterized source but uncharacterized detection devices, a  secret key can be generated with relaxed requirements on the detection and coupling efficiencies, pushing these scenarios now into the domain where experimental realization can be attempted.

\subsection*{Acknowledgments}
D.P., X.F.M. and N.L. have been supported by NSERC under its Discovery Programme, the NSERC-ANR SPG FREQUENCY and the Innovation Platform Quantum Works. Further support has been received by the Ontario Centres of Excellence.   P.v.L and R.W. were supported by the German Research Foundation (Deutsche Forschungsgemeinschaft) through its Emmy Noether Program. Portions of this work were carried out while R.W. was visiting the Institute for Quantum Computing in Waterloo, Canada. He is thankful for the hospitality of the Optical Quantum Communication Theory group and acknowledges financing through the ``Collaborative Student Training in Quantum Information Processing" program.

\bibliography{amplifier}
\end{document}